\documentclass[journal]{IEEEtran}

\usepackage{graphicx}
\graphicspath{ {images/} }
\usepackage{float}
\usepackage{amsmath,amssymb}
\usepackage{booktabs}
\usepackage{array}
\usepackage{soul}
\sethlcolor{lightgray}

\usepackage{tikz}
\usetikzlibrary{calc,shapes,arrows}
\usetikzlibrary{positioning}
\usepackage{tikzpeople}

\usepackage{balance}

\usepackage{cite}

\begin{document}
\title{Primary and Secondary Social Media \\ Source Identification}
\author{Brian~Hosler,~\IEEEmembership{Student~Member,~IEEE}
        ~Matthew~Stamm,~\IEEEmembership{Member,~IEEE}
\thanks{Thank you to Hunter Kippun for your help with data collection.}
\thanks{This material is based upon work supported by the National
Science Foundation under Grant No.1553610.}
}

\markboth{IEEE Signal Processing Letters, 2021}%
{Hosler \MakeLowercase{\textit{et al.}}: Primary and Secondary Social Media Source Identification}

\maketitle
\begin{abstract}

Social networks like Facebook and WhatsApp have enabled users to share images with other users around the world. Along with this has come the rapid spread of misinformation. 
%
One step towards verifying the authenticity of an image is understanding its origin, including it distribution history through social media.  
In this paper, we present a method for tracing the posting history of an image across different social networks. To do this, we propose a 	two-stage deep-learning-based approach, which takes advantage of cascaded fingerprints 	in images left by social networks during uploading. Our proposed system is not reliant upon  metadata or similar easily falsifiable information.  
Through a series of experiments, we show that we are able to outperform existing social media source identification algorithms. 
and identify chains of social networks up to length two with over over 84\% accuracy.

\end{abstract}


\section{Introduction}

The widespread use of social media has resulted in the rapid spread of
information, and with it, misinformation. As a result, a variety of
multimedia forensic techniques have been developed to authenticate
images and videos \cite{milani2012_vidForensOverview,
Stamm2013_IEEEAccess, Bayar2018_OpenSetCamID, bondi2017_CamIDCNN,
Cao2009_DemosCamID}. When determining the credibility of an image, it
is often helpful to consider the credibility of the source. Prior
forensics research aimed at determining an image’s source has largely
fallen into two categories: camera model identification
\cite{Kharrazi2004,
Chen2015_CamID,
ZhaoICIP2016_CompEffCamID,
marra2017_SRMcamID,
Thai2014_noiseCamID,
Milani2014_DemosID,
Junior2019_indepth} and device identification \cite{Lukas2006, Liu2010, Bartlow2009,
Chen2008, Li2010, Costa2014, Kang2012}.
Several of these techniques have
been adapted to operate on video as well
\cite{milani2012_vidForensOverview, Kurosawa1999_CCDfingerprint,
Hosler2019, chen2007_vidPRNU, Mayer_openset}.

One important aspect of image source identification that has received
comparably little attention is determining an image's distribution
channel; specifically determining which, if any, social network have
been used to share an image. The posting history of images is of
particular concern in the case of misinformation campaigns, which are
typically launched over social media. In such campaigns, it is common
for an image to be posted along with text which provides false context,
such as claims of violent rioters at a peaceful protest. In such cases,
knowing that the picture has been reposted, and does not belong to the
purported author, can disprove the falsely supplied context.

When images are uploaded to a social media network, the hosting platform
will often apply a series of processing operations to the image. The
specific operations are proprietary, and therefore unknown, but include
JPEG re-compression and re-sampling(i.e. resizing) as methods for
reducing the size of the image to be stored by the platform. 
Processing operations such as these have been known to leave unique
fingerprints; when operations are cascaded, fingerprints are
likewise cascaded~\cite{Stamm2013_wifs, Bayar2018TowardsOO, conotter2015}.
However when fingerprints are cascaded, older fingerprints are degraded
or obscured. This means that reposting an image to a second social media
platform can obscure the traces left by the original platform. Unless
accounted for, this will result in decreased accuracy when
determining a reposted image's original network.

To date, comparatively little work has been done toward identifying an
image’s posting history.  Amerini et al. used traces of double JPEG
compression to identify which social media site an image was posted
to~\cite{Amerini_DCTtrees}. This technique, however, is only designed to
operate on images posted to a single social network. 
The authors refined this by developing a CNN-based approach capable of
operating on images that may have been uploaded to multiple
networks~\cite{Amerini_DCTCNN}. This system, however, can only identify
the last social network an image is posted to.  Recently, Phan et al.
introduced a system to identify re-posted images, and the social
networks they were reposted from~\cite{Phan2019}. To do this, the
authors used multi-JPEG compression traces, as well as image-wide
features like quantization table coefficients, image dimensions, and
lossless coding parameters.

While prior work provides important first steps toward robust social
media source identification, it still has several shortcomings. The
reliance on easily mutable features such as image dimensions and the
ordering of coding tables makes  Phan et al.'s system unreliable when
presented with unseen image sizes and lossless reordering of the JPEG
image file.

These features can easily be altered, even unintentionally, and without
these features, the system's accuracy is reduced by as much as 30\%.
Therefore, when attempting to determine the posting history of an image,
we must rely on intrinsic or embedded features, as opposed to those
easily altered features such as file structure. Furthermore, if these
intrinsic features could be identified from several smaller regions of
an image, a system could be developed to identify inconsistencies within
the image, offering information as to its authenticity.

In light of this we would like to design a system that can accurately
determine the 
posting history of an image patch, that is not
reliant on unreliable features. Future adaptation of this system could
then localize inconsistencies in the posting history within an image.

In this paper, we propose a patch-based system for tracing the social
network posting history of an image. Unlike the current state of the
art, our system does not rely on falsifiable metadata or non-local
features such as image dimensions. We will show that our system is able
to outperform existing techniques that do not make use of these features.

\section{System Architecture} To perform social media source
identification, we propose a system which leverages both deep learning
techniques and traditional forensic fingerprints. Our proposed system
comprises two stages. The first stage determines the social media
network which most recently hosted an image. This 
is leveraged in the second stage to determine the previous host
network, if any.

This two-stage architecture offers benefits over using a single
classifier. It has been shown that the traces left by one editing
operation can be obscured or overwritten by another editing operation
\cite{Stamm2013_wifs, Bayar2018TowardsOO, conotter2015, Boroumand2017}.
The way in which the original traces are obscured is dependent on the
editing operation.
By using a two-stage system, and first identifying the latest social
network, we can use more specialized classifiers to discriminate between
the potential previous networks.

\begin{figure}
	\centering
	\includegraphics[width=.5\textwidth]{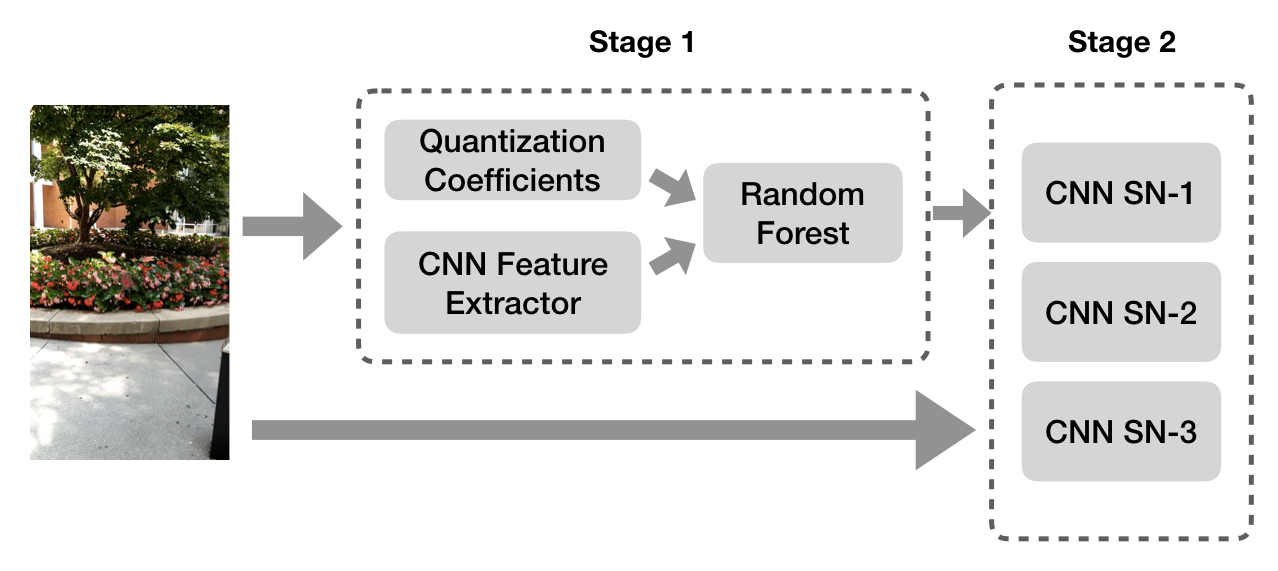}
	\caption{Proposed system architecture. The output of Stage 1 is used
to determine the appropriate Stage 2 classifier.}
	\label{fig:sysarch}
	\vspace{-1em}
\end{figure}

\vspace{-1em}
\subsection{Stage 1}
The first stage of our system determines 
the social media
platform that most recently hosted an image. This stage uses the
deep features of a CNN
trained to discriminate between social networks, as features for
classification. These deep features, as well as compression
parameters of the image, are fed to a random forest classifier. On the
basis of these features, the classifier is able to make highly accurate
decisions.

In order to learn discriminating features, we first train a CNN to
classify images based on their social media of origin. We then remove
the decision function of the network, and use the output neuron
activations as a feature vector.
We use a CNN architecture comprising a single convolutional layer,
followed by four convolutional blocks (Convolution, batch normalization,
tanh activation and max pooling), then two full-connected blocks (matrix
multiplication, activation), and an output layer where each neuron
corresponds to a single class. The first convolutional layer contains 6
filters, each of size $3 \times 3$. The four convolutional blocks have 96 $7
\times 7$ filters, then 64 $5 \times 5$ filters, 64 $5 \times 5$
filters, and finally 128 filters of size $1 \times 1$.
The full-connected layers each have 200
neurons~\cite{BayarEI2017_DesignCNN}.
Instead of designating one output neuron for each social media
processing chain, we assign one neuron for each social media platform.
For example, the Google Photos class encompasses all images which have
been uploaded to Facebook or WhatsApp before being posted to Google
Photos, as well as images which were only posted to Google Photos.

Because errors made in the first stage will propagate through to
following stage system, it is critical that the classifier used in this
stage is as accurate as possible. In order to increase the accuracy of
our first stage, we supplement the features learned by the CNN with
features extracted from the image's encoding process. We observed that
all images remained JPEG-compressed when downloaded from a social
network. However the parameters of this compression, specifically the
quantization table corresponding to the luma channel, were often
changed.

To leverage this change, we extract the first 9 AC coefficients of the
luma-channel quantization table stored in the image.
These features have been shown to be useful for other forensic
tasks~\cite{Kee2011_JPEGheader}.
We use these 
coefficients, and the 
activations of our CNN, as the basis for classification.
For an image patch $X$, our Stage 1 classification, $C$, is given by,
\begin{equation}
	C = \mathcal{F}( \text{CNN(X)} || \mathcal{Q}(X)),
\end{equation}
Where $\text{CNN}(X)$ is the deep features obtained by the CNN,
$\mathcal{Q}(X)$ indicates the quantization table coefficients of the
image, and $||$ is the concatenation of the two vectors. $\mathcal{F}$
is the random forest classifier.

Intuitively, our Stage 1 classification is attained by first
concatenating the output neuron activations of the CNN with the selected
quantization table coefficients. This feature vector is then fed to a
random forest classifier, which is able to achieve a higher accuracy
than the CNN alone.


\subsection{Stage 2}
The second stage of our system determines the
previous social media network of an image, given an image and that
image's latest social network. This stage contains one classifier for
each social network class in the first stage. For the classifiers in
this stage, we use the same CNN architecture as the feature extractor in
the first stage.

Each classifier in this stage is predicated on the input image's latest
social media network and attempts to discern the images previous social
network. To do this, we train each classifier only on images which share
a latest social media network, and designate one output neuron for each
candidate previous social media network. For example, one classifier
will take as input, an image which Stage 1 identified as Google Photos,
and will have output neurons corresponding to None, Facebook(HQ),
Facebook(LQ), and WhatsApp. If the maximum activation of this CNN
corresponds to Facebook(HQ), then the image is classified as having been
uploaded to Facebook(HQ), then downloaded and posted to Google Photos.

This same structure is replicated for every other social network.
Therefore, for any social media source identifiable by the first stage
of our system, there is a corresponding second-stage classifier to
determine the previous social network to host the image in question.
Figure \ref{fig:sysarch} illustrates the whole system architecture.
As we will show,
this system outperforms the current
state of the art system which considers each possible permutation of
social networks concurrently.

\section{Experimental Results}

\subsection{Dataset}
\label{ssec:dataset}
To experimentally verify our system, we created a dataset of images with
11 unique social media upload chains, with a maximum of two social
networks in a chain. In order to create this dataset, we used a subset
of images from the VISION \cite{shullani2017vision} dataset. The VISION
dataset contains natural(NAT) images, as well as copies of those images
uploaded to each of Facebook with the high quality setting (FBH),
Facebook with the low-quality setting (FBL), and WhatsApp (WA). For our
dataset, we took 1219 natural images, and all social media edited copies
from five different camera models of the VISION database. We uploaded
additional copies of the natural images to Google Photos (GOG). This
resulted in four social network processing chains of length one (natural
images uploaded to Facebook-HQ, Facebook-LQ, WhatsApp, or Google
Photos), and one chain of length zero (natural images).

Next, three processing chains of length two were created by reposting
all images with a processing chain of length one to Google Photos,
excluding those which had already been posted to Google Photos. Three
more chains of length two were created by reposting images with a
processing chain of length one to Facebook-HQ, except for those which
had already been posted to Facebook-HQ.
Due to resource and time constraints, the other length-two processing
chains could not be investigated.
This resulted in nearly 13,500 images from 11 unique social media
processing chains.

Images were given both a primary and a secondary label. The primary
label, an upper-case code from \{NAT, FBH, FBL, WA, GOG\} corresponds to
the last social media that an image was uploaded to, which the first
stage of our system is built to identify. An image's secondary label, a
lower-case code from one of \{nat, fbh, fbl, wa, gog\} corresponds to
the social media platform that an image was uploaded to before the
latest one. For example, an image with the label ``fblGOG" was first
uploaded to Facebook-LQ, then uploaded to Google Photos. An image
labeled ``natGOG" was only uploaded to Google Photos and no other
platforms.

\subsection{Stage 1}
To train our Stage 1 classifier, we first trained our feature extraction
CNN. Our dataset was divided into training and testing images by
randomly selecting 10\% of the images from each class for testing, and
leaving the rest for training. From
the training images, we extracted 960,000 image patches of size $64
\times 64$ from images with each of the five primary labels. From the
testing images, 96,000 patches were extracted for each class. We then
trained our CNN feature extractor using stochastic gradient descent with
an initial learning rate of $0.001$, that was halved every 3 epochs for
50 epochs.
We also extracted 240,000 training patches and 24,000 testing
patches of size $128 \times 128$, as well as 60,000 training and 6,000
testing patches of size $256 \times 256$. These datasets were used to
train two additional feature extractors using the same training
procedure.

\begin{table}
\centering
	\caption{Accuracy of classifiers trained to identify
	patch's latest social media network.}
	\label{tab:primary}
\vspace{1.5mm}
	\begin{tabular}{@{}lccc@{}}
	\toprule
		Patch Size&64&128&256\\
	\midrule
		Amerini et al. \cite{Amerini_DCTCNN}&78.1\%&87.7\%&91.7\%\\
		Proposed CNN &83.53\% &92.67\% &86.41\%\\
		Proposed Stage 1&\textbf{93.92\%} &\textbf{94.50\%} &\textbf{94.75\%}\\
	\bottomrule
	\end{tabular}
\vspace{-4mm}
\end{table}

After each CNN was trained, we concatenated the output activations of
each training patch with the first 9 AC quantization table coefficients.
These new feature vectors were then used to train a random forest
classifier. The classifier's accuracy was then scored using the patches
from the testing set, which were not used to train the CNN or the extra
trees classifier.

Table \ref{tab:primary} compares the performance of our classifier at
each patch size with the patch-based classifier proposed in
\cite{Amerini_DCTCNN}, which we implemented and trained.
From Table \ref{tab:primary} we can see that our Stage 1 classifier out-performs
existing work at all patch sizes. Additionally, our classifier
suffers less from the use of small patch sizes than the
existing classifier. The accuracy of our classifier operating on $64
\times 64$ patches is 93.92\%, while the classifier operating on $256
\times 256$ patches is 94.75\%, less than a 1\% decrease. The classifier
in \cite{Amerini_DCTCNN} suffers a 13\% decrease in accuracy between the
smallest and largest patch sizes. These results show that our
classifier architecture is better suited for discriminating 
different social networks.

\textbf{VSMUD Dataset}\\*
For comparison, we trained and evaluated our first stage using the VSMUD
dataset \cite{Phan2019}, which was used to train the current
state-of-the-art system at a comparable patch size.
From this dataset we have selected all images belonging to set C2,
which contains all images with a social network processing chain of
length two or less,
excluding those which have been uploaded to the same platform twice,
resulting in a total of 4,590 images. We labeled these images based
on their latest social media network, resulting in  four classes: FB,
FL, TW, and Original. For example, the TW class includes all images
which were posted directly to Twitter, as well as images which were
posted to Facebook or Flicker, then reposted to Twitter.

These images were divided into a training set and a testing set, from
which $64 \times 64$ patches were extracted using the same process
described in Section \ref{ssec:dataset}. These patches were then used to
train and evaluate our Stage 1 classifier, using the same procedure.
Table \ref{tab:vsmud} compares the accuracy of our proposed Stage 1
classifier to the accuracy of the PCNN classifier reported in
\cite{Phan2019}, which was trained on the same dataset.

Table \ref{tab:vsmud} shows that on this dataset, our classifier attains
an accuracy of 98.3\%. The previous state of the art classifier, PCNN,
correctly classifies only 92.3\% of image patches. We note that, using
our CNN alone as a classifier, we can achieve a higher accuracy than
the PCNN. These results confirm the previous experiment which showed
that our classifier out-performs the existing work.

\begin{table}
\centering
	\caption{Accuracy of classifiers trained using the
	VSMUD \cite{Phan2019} dataset with a patch size of 64.}
	\label{tab:vsmud}
\vspace{1.5mm}
	\begin{tabular}{@{}lccc@{}}
	\toprule
		Classifier&PCNN\cite{Phan2019}&Proposed CNN&Proposed Stage 1\\
	\midrule
		Accuracy & 92.3\% & 92.8\% & \textbf{98.3\%}\\
	\bottomrule
	\end{tabular}
\vspace{-4mm}
\end{table}

\subsection{Stage 2}

Our system's second stage contains one CNN classifier for each class in
Stage 1.
To train Stage 2, we trained each CNN
individually, using images that shared a primary label, i.e. images
whose processing chain all end with the same social network. From the images
in the training set with GOG as their primary label, we extracted 10,000
training patches and 1,000 testing patches of size $256 \times 256$ from
images with each secondary label. The same was done for images with FBH
as their primary label. From each secondary class, 40,000 training
patches and 4,000 testing patches of size $128 \times 128$ were
extracted for both GOG and FBH images. For a patch size of $64 \times
64$, 160,000 training patches and 16,000 testing patches were selected
from each class.

One classifier was trained using each training set and evaluated with
the corresponding testing set. Stage 2 classifiers were trained with an
initial learning rate of $0.005$, which was decayed by a factor of 0.7
every 3 epochs for 60 epochs. Table~\ref{tab:double} shows the accuracy
of each classifier

\begin{table}
	\centering
	\caption{Stage 2 classifier accuracies.}
	\label{tab:double}
	\begin{tabular}{cccc}
		\toprule
		Primary Class & \multicolumn{3}{c}{Patch Size} \\
		& $256\times 256$ & $128\times 128$ & $64\times 64$ \\
		\midrule
		GOG & 91.4\% & 88.2\% & 85.3\% \\
		FBH & 90.6\% & 89.8\% & 84.0\% \\
		\bottomrule
	\end{tabular}
	\vspace{-1em}
\end{table}

As shown in Table \ref{tab:double}, both Stage 2 classifiers achieve an
average accuracy of 91\% with a patch size of $256 \times 256$. Our
classifiers are highly accurate at all tested patch sizes, with a
minimum of 84\% of patches correctly identified. These results suggest
that, if the primary label is known, our Stage~2 classifier can reliably
determine the previous social media to which an image was uploaded.

\subsection{System Level}

To evaluate the performance of our system as a whole, we selected 16,000
$64\times64$ testing patches from each of the 11 unique social media
processing chains. We then used our pre-trained Stage~1 classifier to
determine the primary label of each patch. 
Patches that were identified as having a GOG or FBH primary label were
sent to their respective Stage~2 classifier to examine for evidence of
re-posting. 
%
We also trained and evaluated Phan et al.'s P-CNN classifier as
described in~\cite{Phan2019}.  For fair comparison, training was done
using $160,000$ training patches per class from our dataset (i.e. the
same amount of training data used for our proposed classifier). We then
tested using $16,000$ testing patches per class from
our dataset. 

\begin{table}
\centering
\caption{Stage 1 classifier confusion matrix by social media path.}
\label{tab:sys1}
\begin{tabular}{r|ccccc}
& \multicolumn{5}{c}{Predicted class}\\
True Path & NAT & GOG & FBH & FBL & WA \\
\midrule
	natNAT & \hl{99.93} &  --  &  --  &  --  & 0.07 \\
	natGOG &  --  & \hl{100.0} &  --  &  --  &  --  \\
	fbhGOG &  --  & \hl{100.0} &  --  &  --  &  --  \\
	fblGOG &  --  & \hl{99.24} & 0.01 & 0.75 &  --  \\
	waGOG  &  --  & \hl{100.0} &  --  &  --  &  --  \\
	natFBH &  --  &  --  & \hl{89.75} & 10.25 &  --  \\
	gogFBH &  --  &  --  & \hl{99.55} & 0.45 &  --  \\
	fblFBH &  --  &  --  & \hl{12.84} & 87.16 &  --  \\
	waFBH  & 1.88 &  --  & \hl{72.11} & 24.14 & 1.88 \\
	natFBL &  --  &  --  & 19.22 & \hl{80.78} &  --  \\
	natWA  &  --  &  --  &  --  &  --  & \hl{100.0} \\
\end{tabular}
	\vspace{-1em}
\end{table}

In this experiment, our proposed system was able to correctly classify
77.09\% of patches.   By contrast, the P-CNN correctly classified 14.4\%
of patches. This is a difference of of over 60 percentage points in
accuracy.  
These results show that our classifier  significantly outperforms Phan
et al.'s P-CNN at performing patch-level social media source
identification.  

Breaking down our results by stage, a more detailed look at our system's
performance can be obtained.  Our Stage 1 classifier was able to
correctly classify 86.76\% of testing patches. Our Stage 2 classifiers
achieved accuracies of 78.4\% for the FBH classifier and 89.3\% for the
GOG classifier. This results in an overall system accuracy of 77.09\%.
The difference in accuracy between our Stage 1 classifier in this
experiment and in the results presented in Table \ref{tab:sys1} is due
to an intentional imbalance in the classes to represent each
processing chain equally.

Table \ref{tab:sys1} shows the confusion matrix of our Stage 1
classifier in this experiment. From Table \ref{tab:sys1} we can see that
the Natural, Google Photos and WhatsApp primary labels are easily
distinguished from each other. The Facebook-HQ and Facebook-LQ classes
are easily confused, causing a slight decrease in accuracy. However,
this decrease is much less severe than that of the P-CNN classifier.

These results show that our multi-stage classifier approach is better
able to differentiate social network posting histories.
This is especially significant when we consider that our system is
vulnerable to errors in Stage 1 being cascaded into Stage 2. This
suggests that our proposed system is able to take advantage of cascaded
fingerprints, instead of merely discriminate between them.

We note that the accuracy that the P-CNN achieved in this experiment was
significantly lower than the accuracy reported by Phan et al. in their
original work.   While the original P-CNN classifier was not made
publicly available, we confirmed our implementation through
correspondence with the author.
There are several possible reasons for this decreased performance.
One factor we believe contributed to this difference is the difference
between class definitions, particularly the inclusion of both FBH and
FBL, as compared to Phan et al.'s singular FB. We believe that the PCNN
architecture is particularly sensitive to JPEG quantization, and while
FBH and FBL are very different processing chains, they have very similar
JPEG compression.
Our experiments also considered natural images whereas Phan et al.'s did
not.
Additionally, it is possible that our larger network could better take
advantage of the larger volume of training data in our experiments than
the much lighter weight P-CNN network.  
These results further demonstrate the advantage of our proposed approach.

\section{Conclusion}


In this paper, we propose a multistage system for tracing the social network posting history of an image. Our proposed system does not rely on mutable  non-local information such as the ordering of JPEG headers or the dimensions of an image. 
Instead, it utilizes a CNN to learn traces left by social networks from small patches within an image.  
Through a series of experiments, we demonstrated that our classifier achieves an overall accuracy of over 84\%, and outperforms comparable existing work.

\balance
\bibliographystyle{IEEEbib}
\footnotesize
\bibliography{sourceID}

\begin{thebibliography}{10}

\bibitem{milani2012_vidForensOverview}
S.~Milani, M.~Fontani, P.~Bestagini, M.~Barni, A.~Piva, M.~Tagliasacchi, and
  S.~Tubaro,
\newblock ``An overview on video forensics,''
\newblock {\em APSIPA Transactions on Signal and Information Processing}, vol.
  1, 2012.

\bibitem{Stamm2013_IEEEAccess}
M.~C. Stamm, M.~Wu, and K.~J.~R. Liu,
\newblock ``Information forensics: An overview of the first decade,''
\newblock {\em IEEE Access}, vol. 1, pp. 167--200, 2013.

\bibitem{Bayar2018_OpenSetCamID}
B.~Bayar and M.~C. Stamm,
\newblock ``Towards open set camera model identification using a deep learning
  framework,''
\newblock in {\em IEEE International Conference on Acoustics, Speech, and
  Signal Processing (ICASSP)}, Calgary, Canada, Apr. 2018.

\bibitem{bondi2017_CamIDCNN}
L.~Bondi, L.~Baroffio, D.~Güera, P.~Bestagini, E.~J. Delp, and S.~Tubaro,
\newblock ``First steps toward camera model identification with convolutional
  neural networks,''
\newblock {\em IEEE Signal Processing Letters}, vol. 24, no. 3, pp. 259--263,
  Mar. 2017.

\bibitem{Cao2009_DemosCamID}
H.~Cao and A.~C. Kot,
\newblock ``Accurate detection of demosaicing regularity for digital image
  forensics,''
\newblock {\em IEEE Transactions on Information Forensics and Security}, vol.
  4, no. 4, pp. 899--910, Dec. 2009.

\bibitem{Kharrazi2004}
M.~{Kharrazi}, H.~T. {Sencar}, and N.~{Memon},
\newblock ``Blind source camera identification,''
\newblock in {\em 2004 International Conference on Image Processing, 2004. ICIP
  '04.}, Oct 2004, vol.~1, pp. 709--712 Vol. 1.

\bibitem{Chen2015_CamID}
C.~Chen and M.~C. Stamm,
\newblock ``Camera model identification framework using an ensemble of
  demosaicing features,''
\newblock in {\em IEEE International Workshop on Information Forensics and
  Security (WIFS)}, Nov. 2015, pp. 1--6.

\bibitem{ZhaoICIP2016_CompEffCamID}
X.~Zhao and M.~C. Stamm,
\newblock ``Computationally efficient demosaicing filter estimation for
  forensic camera model identification,''
\newblock in {\em IEEE International Conference on Image Processing (ICIP)},
  Sept. 2016, pp. 151--155.

\bibitem{marra2017_SRMcamID}
F.~Marra, G.~Poggi, C.~Sansone, and L.~Verdoliva,
\newblock ``A study of co-occurrence based local features for camera model
  identification,''
\newblock {\em Multimedia Tools and Applications}, vol. 76, no. 4, pp.
  4765--4781, 2017.

\bibitem{Thai2014_noiseCamID}
T.~H. Thai, R.~Cogranne, and F.~Retraint,
\newblock ``Camera model identification based on the heteroscedastic noise
  model,''
\newblock {\em IEEE Transactions on Image Processing}, vol. 23, no. 1, pp.
  250--263, Jan. 2014.

\bibitem{Milani2014_DemosID}
S.~Milani, P.~Bestagini, M.~Tagliasacchi, and S.~Tubaro,
\newblock ``Demosaicing strategy identification via eigenalgorithms,''
\newblock in {\em IEEE International Conference on Acoustics, Speech and Signal
  Processing (ICASSP)}, May 2014, pp. 2659--2663.

\bibitem{Junior2019_indepth}
P.~R. {Mendes J\'unior}, L.~{Bondi}, P.~{Bestagini}, S.~{Tubaro}, and
  A.~{Rocha},
\newblock ``An in-depth study on open-set camera model identification,''
\newblock {\em IEEE Access}, vol. 7, pp. 180713--180726, 2019.

\bibitem{Lukas2006}
J.~{Lukas}, J.~{Fridrich}, and M.~{Goljan},
\newblock ``Digital camera identification from sensor pattern noise,''
\newblock {\em IEEE Transactions on Information Forensics and Security}, vol.
  1, no. 2, pp. 205--214, June 2006.

\bibitem{Liu2010}
B.~{Liu}, Y.~{Hu}, and H.~{Lee},
\newblock ``Source camera identification from significant noise residual
  regions,''
\newblock in {\em 2010 IEEE International Conference on Image Processing}, Sep.
  2010, pp. 1749--1752.

\bibitem{Bartlow2009}
N.~{Bartlow}, N.~{Kalka}, B.~{Cukic}, and A.~{Ross},
\newblock ``Identifying sensors from fingerprint images,''
\newblock in {\em 2009 IEEE Computer Society Conference on Computer Vision and
  Pattern Recognition Workshops}, June 2009, pp. 78--84.

\bibitem{Chen2008}
M.~{Chen}, J.~{Fridrich}, M.~{Goljan}, and J.~{Lukas},
\newblock ``Determining image origin and integrity using sensor noise,''
\newblock {\em IEEE Transactions on Information Forensics and Security}, vol.
  3, no. 1, pp. 74--90, 2008.

\bibitem{Li2010}
C.~{Li},
\newblock ``Source camera identification using enhanced sensor pattern noise,''
\newblock {\em IEEE Transactions on Information Forensics and Security}, vol.
  5, no. 2, pp. 280--287, 2010.

\bibitem{Costa2014}
Filipe {de O. Costa}, Ewerton Silva, Michael Eckmann, Walter~J. Scheirer, and
  Anderson Rocha,
\newblock ``Open set source camera attribution and device linking,''
\newblock {\em Pattern Recognition Letters}, vol. 39, pp. 92--101, 2014,
\newblock Advances in Pattern Recognition and Computer Vision.

\bibitem{Kang2012}
X.~{Kang}, Y.~{Li}, Z.~{Qu}, and J.~{Huang},
\newblock ``Enhancing source camera identification performance with a camera
  reference phase sensor pattern noise,''
\newblock {\em IEEE Transactions on Information Forensics and Security}, vol.
  7, no. 2, pp. 393--402, 2012.

\bibitem{Kurosawa1999_CCDfingerprint}
K.~Kurosawa, K.~Kuroki, and N.~Saitoh,
\newblock ``Ccd fingerprint method-identification of a video camera from
  videotaped images,''
\newblock in {\em International Conference on Image Processing}, Oct. 1999,
  vol.~3, pp. 537--540.

\bibitem{Hosler2019}
B.~{Hosler}, O.~{Mayer}, B.~{Bayar}, X.~{Zhao}, C.~{Chen}, J.~A. {Shackleford},
  and M.~C. {Stamm},
\newblock ``A video camera model identification system using deep learning and
  fusion,''
\newblock in {\em ICASSP 2019 - 2019 IEEE International Conference on
  Acoustics, Speech and Signal Processing (ICASSP)}, 2019, pp. 8271--8275.

\bibitem{chen2007_vidPRNU}
M.~Chen, J.~Fridrich, M.~Goljan, and J.~Luk{\'a}{\v{s}},
\newblock ``Source digital camcorder identification using sensor photo response
  non-uniformity,''
\newblock in {\em Security, Steganography, and Watermarking of Multimedia
  Contents IX}. International Society for Optics and Photonics, 2007, vol.
  6505, p. 65051G.

\bibitem{Mayer_openset}
O.~{Mayer}, B.~{Hosler}, and M.~C. {Stamm},
\newblock ``Open set video camera model verification,''
\newblock in {\em ICASSP 2020 - 2020 IEEE International Conference on
  Acoustics, Speech and Signal Processing (ICASSP)}, 2020, pp. 2962--2966.

\bibitem{Stamm2013_wifs}
M.~C. {Stamm}, X.~{Chu}, and K.~J.~R. {Liu},
\newblock ``Forensically determining the order of signal processing
  operations,''
\newblock in {\em 2013 IEEE International Workshop on Information Forensics and
  Security (WIFS)}, Nov 2013, pp. 162--167.

\bibitem{Bayar2018TowardsOO}
Belhassen Bayar and Matthew~C. Stamm,
\newblock ``Towards order of processing operations detection in jpeg-compressed
  images with convolutional neural networks,''
\newblock in {\em Media Watermarking, Security, and Forensics}, 2018.

\bibitem{conotter2015}
V.~{Conotter}, P.~{Comesaña}, and F.~{Pérez-González},
\newblock ``Forensic detection of processing operator chains: Recovering the
  history of filtered jpeg images,''
\newblock {\em IEEE Transactions on Information Forensics and Security}, vol.
  10, no. 11, pp. 2257--2269, Nov 2015.

\bibitem{Amerini_DCTtrees}
Irene Amerini, Roberto Caldelli, Andrea~Del Mastio, Andrea~Di Fuccia, Cristiano
  Molinari, and Anna~Paola Rizzo,
\newblock ``Dealing with video source identification in social networks,''
\newblock {\em Signal Processing: Image Communication}, vol. 57, pp. 1 -- 7,
  2017.

\bibitem{Amerini_DCTCNN}
I.~{Amerini}, T.~{Uricchio}, and R.~{Caldelli},
\newblock ``Tracing images back to their social network of origin: A cnn-based
  approach,''
\newblock in {\em 2017 IEEE Workshop on Information Forensics and Security
  (WIFS)}, Dec 2017, pp. 1--6.

\bibitem{Phan2019}
Q.~{Phan}, G.~{Boato}, R.~{Caldelli}, and I.~{Amerini},
\newblock ``Tracking multiple image sharing on social networks,''
\newblock in {\em ICASSP 2019 - 2019 IEEE International Conference on
  Acoustics, Speech and Signal Processing (ICASSP)}, May 2019, pp. 8266--8270.

\bibitem{Boroumand2017}
Mehdi Boroumand and Jessica~J. Fridrich,
\newblock ``Scalable processing history detector for jpeg images,''
\newblock in {\em Media Watermarking, Security, and Forensics}, 2017.

\bibitem{BayarEI2017_DesignCNN}
B.~Bayar and M.~C. Stamm,
\newblock ``Design principles of convolutional neural networks for multimedia
  forensics,''
\newblock in {\em IS\&T Symposium on Electronic Imaging (EI) - Media
  Watermarking, Security, and Forensics - Special Session on Deep Learning for
  Multimedia Security}, San Francisco, CA, Feb. 2017, pp. 77--86.

\bibitem{Kee2011_JPEGheader}
E.~Kee, M.~K. Johnson, and H.~Farid,
\newblock ``Digital image authentication from jpeg headers,''
\newblock {\em IEEE Transactions on Information Forensics and Security}, vol.
  6, no. 3, pp. 1066--1075, Sept. 2011.

\bibitem{shullani2017vision}
D.~Shullani, M.~Fontani, M.~Iuliani, O.~Al~Shaya, and A.~Piva,
\newblock ``Vision: a video and image dataset for source identification,''
\newblock {\em EURASIP Journal on Information Security}, vol. 2017, no. 1, pp.
  15, 2017.

\end{thebibliography}

\end{document}